\documentclass[fleqn,10pt]{wlscirep}
\title{Symmetry warrants rational cooperation by co-action in Social 
Dilemmas}

\author[1,*]{V. Sasidevan}
\author[2]{Sitabhra Sinha}
\affil[1]{The Institute of Mathematical Sciences, CIT Campus, Taramani, Chennai 600113, India.}
\affil[2]{The Institute of Mathematical Sciences, CIT Campus, Taramani, Chennai 600113, India.}

\affil[*]{sasidevan@imsc.res.in, sasidevan@gmail.com}

\begin{abstract}
Is it rational for selfish individuals to cooperate? 
The conventional answer based on analysis of games such as the Prisoners
Dilemma (PD) is that it is not, even though mutual cooperation results
in a better outcome for all.
This incompatibility
between individual rationality and collective benefit lies at the
heart of questions about the evolution of cooperation,
as illustrated by PD and similar games.
Here, we argue that this apparent incompatibility is due to an
inconsistency in the standard Nash framework for analyzing non-cooperative
games and propose a new paradigm, that of the
co-action equilibrium. 
As in the Nash solution, agents know that others are just as rational
as them and taking this into account leads them
to realize that others will independently adopt the same strategy,
in contrast to the idea of unilateral deviation central to
Nash equilibrium thinking.
Co-action equilibrium results in better collective
outcomes for games representing social dilemmas, with relatively
``nicer'' strategies being chosen by rational selfish
individuals.
In particular,
the dilemma of PD gets resolved within this framework, suggesting
that cooperation can evolve in nature as the rational outcome even for
selfish agents, without having to take recourse to additional
mechanisms for promoting it.
\end{abstract}
\begin{document}

\flushbottom
\maketitle
% * <john.hammersley@gmail.com> 2015-02-09T12:07:31.197Z:
%
%  Click the title above to edit the author information and abstract
%
\thispagestyle{empty}

% \noindent Please note: Abbreviations should be introduced at the first mention in the main text – no abbreviations lists. Suggested structure of main text (not enforced) is provided below.
\section*{Introduction}
\label{sec1}
Strategic interactions
%, modeled by games, 
occur all around us in a multitude 
of forms
between autonomous {\em agents}. These interacting agents could
correspond to individual humans or animals or
even computer algorithms, as
well as, collective entities such as groups, organizations or nations.
Analyzing their interactions in terms of games~\cite{Morgenstern44} 
is a promising approach for understanding the behavior of a wide
variety of socio-economic and biological systems, and finds
applications in fields ranging from economics and political science to
computer
science and evolutionary biology \cite{colman}.
A {\em game} is described by the set of all possible actions by a specified
number of agents, where each possible combination of actions is
associated with a payoff for each agent.
Thus, the payoff received by an agent depends on 
her choice of action, as well as that of others.  
Agents are assumed to be rational and selfish, who want to maximize
their individual payoffs.
In addition,
every agent knows that all agents satisfy these criteria (for a detailed
discussion of these ideas see, e.g., Ref.~\cite{hargreaves}).
Each of these assumptions is crucial in determining the outcome of a
game. 
While they may or may not hold in specific real-life scenarios,
the agent behavior embodied by these assumptions provides a crucial
benchmark for strategic behavior. 

In order to solve a game, i.e., to find the set of actions that
the agents will employ given the structure of the game, one needs
a solution concept
that will form the basis for
strategy selection by the agents.
For  non-cooperative games, where agents choose their actions
independently without communicating with other agents, the 
canonical solution concept employed is that of the Nash equilibrium.
It is defined informally as the set of actions chosen by the agents
where no agent can gain by unilaterally deviating from this
equilibrium~\cite{osborne}. Nash equilibria exist for all
games having a finite number of agents choosing from a finite set of
actions, making it a very general concept that has wide 
applicability~\cite{Nash1950}. Indeed, the concept has been central to
various
attempts at developing quantitative descriptions of socio-economic
phenomena~\cite{Holt2004}.
However, analyzing specific games using the concept of Nash equilibrium can
raise the following issues: (i) A game may have more than one Nash
equilibria and hence, deciding which of these will be adopted by rational
agents is a non-trivial problem~\cite{harsanyi88}.
Additional criteria need to be provided for selecting an equilibrium;
however their success is not always guaranteed~\cite{binmore}.
(ii) The Nash equilibrium of a game may sometimes be inferior
to an alternative choice of actions by the agents in which all the
parties get higher payoff. This gives rise to apparently paradoxical
situations in games representative of social dilemmas,
such as the Prisoner's Dilemma (PD)~\cite{Rapoport1965}, the Traveler's
Dilemma~\cite{basu1,basu2}, etc. For example,
in PD, where each agent has the option to either cooperate with the
other agents or defect, mutual defection is the only Nash equilibrium,
although mutual cooperation will result in higher payoffs for all
agents. Results of experimental realizations of such games also show
deviation from the Nash solutions~\cite{Rapoport1965,andreoni}.
That rational action by individual agents can
result in an undesirable collective outcome for the agents
is a long-standing puzzle~\cite{Kollock98}. In particular, it raises
questions about how cooperation could have evolved and is maintained
in natural populations~\cite{axelrod}.
%and a considerable body of research
%literature exists devoted to understanding it~\cite{Kollock98}.

Here we argue that the genesis of this problem 
can be
traced to a mutual inconsistency between the assumptions underlying
the Nash equilibrium
%%%NEW June 19
for symmetric game situations. 
%%%NEW June 19
One of these assumptions is that each agent is equally capable of
analyzing the game situation and that all of them are aware of this. 
However, it is also assumed that 
agents can make {\em unilateral} deviations in their strategy,
which is used to obtain a dominant strategy
in games like PD.
In other words, each agent
looks {\em only} at the payoff structure of the game and
takes a decision that is independent of {\em how} other agents decide.
This is inconsistent with the earlier assumption because if the agents
are aware that the others are also rational, they should take this
(rational decision-making by the other agents)
into account.
%%%%NEW%%%%
To put it informally, the player will argue that ``if the
other player is like me, then she will be independently
choosing the same strategy (although not necessarily the same action
if it is a mixed strategy) as I, because we are faced with the same
situation.''
%%%%%%%%%%%
In this paper we present
a novel solution paradigm for payoff-symmetric games,
referred to as {\em co-action equilibrium}, that resolves this
inconsistency, building on a concept originally introduced in
the context of minority games~\cite{sasidevan}.
As we shall see, 
the optimal action of rational agents in co-action equilibrium
is markedly different
from Nash equilibrium and leads to better collective outcomes,
solving various social dilemmas such as PD.

The mutual inconsistency between (i) the assumption of players being aware that
all of them are rational and (ii) the possibility of a dominant strategy,
had been earlier pointed out informally in the specific context of PD
- although, to the
best of our knowledge, there have been no attempts to develop a
quantitative framework that addresses this problem.
In what is possibly the earliest statement about the rationality of
cooperation in PD, 
Rapoport~\cite{rapoport1966} had argued that because of the symmetry 
of the game, rational players will choose the same action - and as it
involves a higher payoff, they will always opt for mutual cooperation. 
This argument has been independently put forward by
Hofstadter~\cite{hofstadter} in the context of a $N$-person PD. 
The response of conventional game theory to this line of reasoning, as 
set forth at length by Binmore~\cite{Binmore1994}, centers on the
argument that these approaches crucially rely on constraining the set
of feasible outcomes of the game to the main diagonal of
the payoff matrix, thereby making it effectively a
collective decision-making process~\cite{McMahon2001}. 
As outlined in detail below, 
the co-action approach presented here allows the agents
access to the full set of outcomes in the game matrix and
the solution is obtained without restricting their choices of action.
%is not built upon such
%restrictions and allows the agents
%access to the full set of outcomes in the game matrix.
%same strategies as opposed to same actions
%Moreover, the solution framework is general, applicable
It is also general, applying to all symmetric
non-cooperative games. 
As the theory of strategic interactions is central to the analysis of many 
phenomena across economics, social sciences and
evolutionary biology,
%Re-examining them in light of 
the co-action concept could potentially lead to new insights
across a broad range of
disciplines.

%This inconsistency has been alluded to by others informally,
%e.g., in the context of PD by Hofstadter~\cite{hofstadter}, although, to the
%best of our knowledge, there have been no attempts to develop a
%quantitative framework that addresses this problem. 
%%% Detailed discussion of the similarity idea
%%% see Colman and Pulford 2012, p 44-45
%Talk about Rapoport 1966 - discussed by Davis in 1977.
%Earlier discussion by \cite{davis1977}.
%Rapoport 1967 switches to a different argument - that of metagames due
%to Howard (see Howard 1988)
%Chapter 3 of Ref.~\cite{Binmore1994} attempts to refute perceived
%``fallacies'' underlying the similarity argument for justifying
%cooperation in PD.
%%we got the reference for Binmore from Colman 2003 Behavioral and
%%Brain Sciences
%Essentially Binmore's refutation is based on the fact that all these
%approaches constrain the game to the diagonal of the payoff matrix.
%However the present approach is not built upon such restrictions.
%Distinguish between players having similar actions vs similar
%strategies. 
%%% EXPAND

In this paper we analyze single-stage games with two actions per
agent, where the payoff
structure is unchanged on exchanging the identities of the agents
(payoff symmetry).
We primarily focus on two-person games, with
agents playing the game once (in contrast to repeated
games where agents can interact many times in an iterative
manner)
and analyze in 
detail three well-known instances, viz., PD, Chicken (also referred to as
snow-drift or Hawk-Dove) and Stag Hunt. 
These games model a wide variety of conflict situations in nature 
where cooperation may emerge under certain
circumstances~\cite{Archetti2012}.
We describe the co-action solution for
these games which, in general, leads to ``nicer'' strategies being
selected by the agents compared to the Nash solution. 
For example, the co-action equilibrium in PD
corresponds to full cooperation among agents at lower values of
temptation to defect, while for higher temptation each agent employs a
probabilistic  strategy.
Thus, co-action typically results in more
globally efficient outcomes,
reconciling the apparent conflict between individual rationality and
collective benefit. 
Further, the co-action equilibrium is unique and therefore, agents are
not faced with the problem of equilibrium selection.
The concept can be extended to
other scenarios, such as, symmetric games involving several players, or
even non-symmetric games when agents can be grouped into clusters
with symmetry holding within each.
In fact, the latter case can be seen as defining a new class of games
between players, where each ``player'' represents a group of agents
who independently choose the same strategy.

\section*{The Co-action equilibrium}
\label{sec2}
\begin{figure}
\begin{center}
\includegraphics[scale = .4]{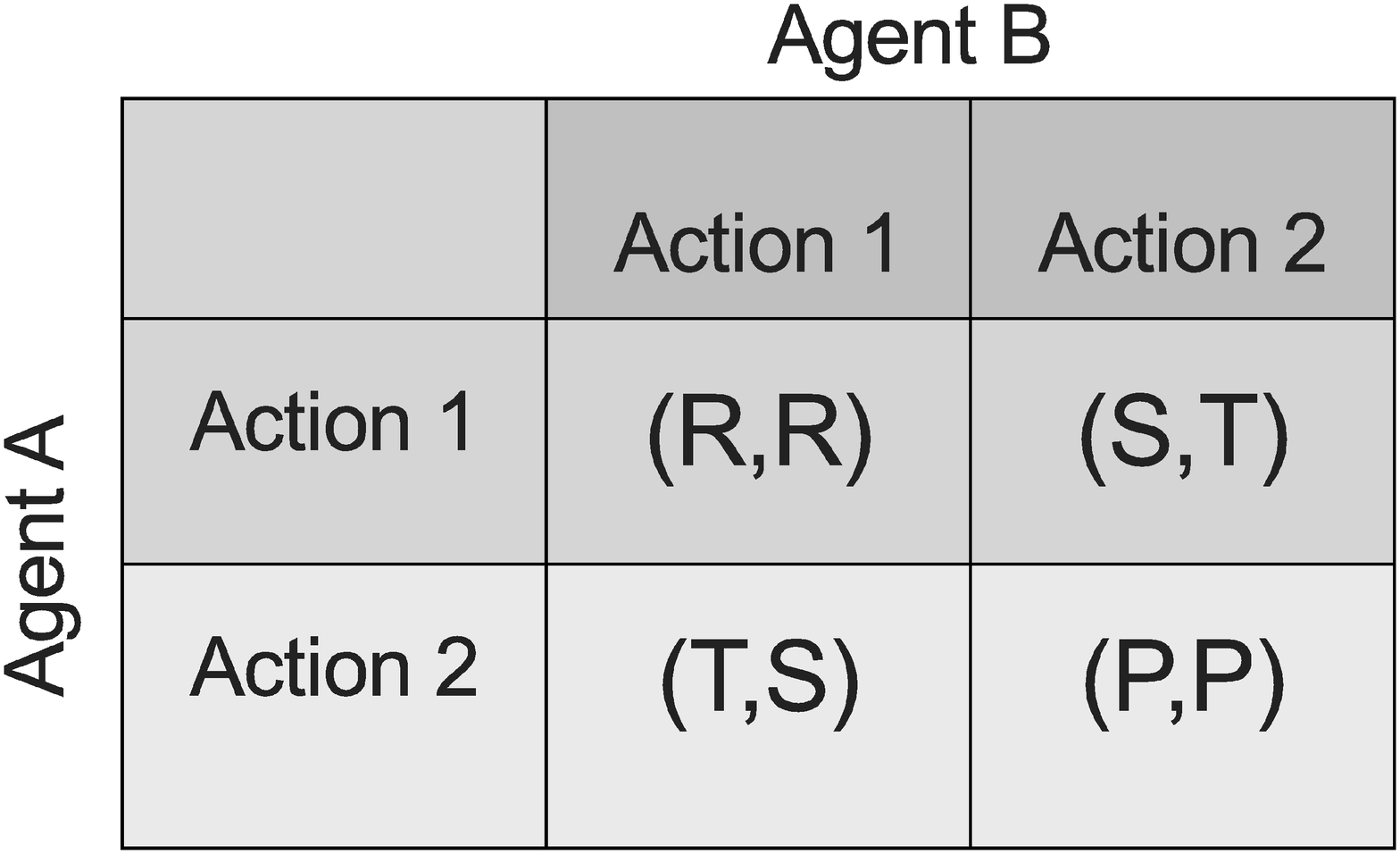}
\end{center}
\caption{A generic representation of the payoff matrix for a
two-person
symmetric game where each agent has two actions available to her.
For
each pair of actions, the first entry in each payoff pair belongs to
Agent {\em A} while the second belongs to Agent {\em B}. Different
games discussed in the text, such as PD, Chicken and Stag-hunt, are
defined in terms of different hierarchical relations among the
elements $T$, $R$, $P$ and $S$.}
\label{fig_1}
\end{figure}

To describe the co-action solution concept, we 
consider the general case of a payoff-symmetric, two-person game where
each agent (say, {\em A} and {\em B}) has
two possible actions (Action~1 and Action~2) available to her.
Each agent receives a payoff corresponding to the pair of choices made
by them. If both agents choose the same option, Action 1 (or 2),
each receives the payoff $R$ (or $P$, respectively), while if they
opt for different choices, the agent choosing Action 1 receives payoff 
$S$ while the other receives $T$.
Thus, the game can be represented by a payoff matrix that
specifies all  possible outcomes (Fig.~\ref{fig_1}). 
An agent may employ a mixed strategy,
in which she randomly selects her options, choosing Action~1 with
some probability $p$ (say) and Action~2 with probability $(1-p)$. 
A pure strategy corresponds to $p$ being
either 0 or 1. 
A Nash equilibrium for a game can be in pure strategies or in mixed
strategies. As noted earlier, a given game may have more than one Nash
equilibrium, possibly involving mixed strategies. Assuming that
agent {\em A} ({\em B}) chooses Action~1 with probability $p_1$ ($p_2$) and
Action~2 with probability $1-p_1$ ($1- p_2$, respectively), their
expected payoffs are,
\begin{align}
\nonumber W_A &= p_1(p_2(R+P-T-S)+S-P) + p_2(T-P) + P,\\
W_B &= p_2(p_1(R+P-T-S)+S-P) + p_1(T-P) + P.
\label{nash0}
\end{align}
The symmetry of the game is reflected in the fact that $W_A$ and $W_B$
are interchanged on exchanging $p_1$ with $p_2$. It is easily seen
that if a mixed strategy Nash equilibrium exists, 
%the optimal strategies 
it is the same for both agents and given by the
probabilities
\begin{equation}
p_1^* = p_2^* =  \dfrac{P-S}{(R+P-T-S)}.
\label{nash}
\end{equation}

The Nash solution assumes that all agents are rational and
that each agent knows the 
planned equilibrium strategies of the other agents. Furthermore, a unilateral
deviation in strategy by one of them will not change the strategy
choice of others (who are assumed to be just as rational as the
one who deviated!). 
This is implicit in Eq.~\ref{nash0} where each agent maximizes her
payoff independent of the strategy of the other agent.
In other words, while making a choice the agents do not take into 
account the fact that the
other agents (who are assumed to have identical capabilities) 
are also deciding simultaneously on their choice and that they
are all aware of this.
Although this latter assumption is deeply embedded
in standard game theory,  it is
inconsistent with the assumption that every agent is aware that all
other agents are just as rational as them.
By contrast, in the co-action concept, by virtue of the symmetry of the game, 
each agent
will argue that whatever complicated processes she employs in
arriving at the optimal decision, the other agents will choose the
same strategy as they 
have the same information and capabilities. 
It is important to note that this does not require any communication
between the agents nor does it invoke the existence of trust or other
extraneous concepts. Rather, it arises from the fact that
both agents are equally rational and being in a
symmetric situation,  will reach the same conclusion about the choice
of strategy; moreover, {\it they realize and consider this in making
their decision}.
It is important to note that the co-action
concept does not imply that both agents will necessarily end up 
choosing the same action.
For instance, 
the co-action solution for the single-stage PD is not to always 
cooperate -
%as suggested by Rapoport and Hofstadter's
%argument~\cite{rapoport1966,hofstadter}) 
which distinguishes the present approach from the earlier arguments of
Rapoport~\cite{rapoport1966} and Hofstadter~\cite{hofstadter} where all
agents always choose the same action -
but to resort to a
mixed strategy when the temptation to
defect is sufficiently high.

In the co-action concept, each agent maximizes her payoff assuming
that all other agents in a symmetric situation will be making the same
decision. Formally this amounts to optimizing the expected
payoff functions of each of the two agents, which in this case are identical:
\begin{equation}
W_{A,B} = W = p^2 (R+P-T-S) + p (T+S-2P) + P.
 \label{co-action_payoff}
\end{equation}
Here $p$ is the probability with which each of the
agents {\em A} and {\em B} chooses Action 1.
Under the co-action concept, the equilibrium strategy $p^*$ of the agents is
obtained by maximizing $W$ with respect to $p \in [0,1]$.
If the maximum of function $W$ in $[0,1]$ occurs at one of the ends
(i.e., $p = 0$ or $1$), it results in a pure strategy co-action
equilibrium. However, if $W$ has a maximum inside $(0,1)$ then the
co-action equilibrium is a non-trivial mixed strategy, viz.,
\begin{equation}
 p^* = \dfrac{2P-(T+S)}{2(R+P-T-S)}.
\label{eq3}
\end{equation}
The existence of the co-action equilibrium for
all symmetric games is guaranteed from the smoothness of polynomial
functions such as Eq.~\ref{co-action_payoff}.
Also, unlike the Nash equilibrium, the co-action
equilibrium is unique and thus, for a given symmetric game there is no
ambiguity about the optimal choice of action for the agents.

\section*{Case studies}
\label{sec3}
Having described the concept of co-action
equilibrium, we will now apply it
to three well-known two-person symmetric games, illustrating in each
case the differences between the co-action and Nash equilibria. Each of these
games is defined in terms of a specific hierarchical relationship
between the payoffs $R$, $S$, $T$ and $P$ (using the terminology of
the payoff matrix shown in Fig.~\ref{fig_1}).

\subsection*{Prisoner's Dilemma}
\label{pd}
PD is one of the most well-studied games in the
literature of strategic choices in social sciences and evolutionary
biology~\cite{rapoport,nowak2004}. 
It is the canonical paradigm for analyzing the problems
associated with 
evolution of cooperation among selfish
individuals~\cite{axelrod}.  
The game represents a
strategic interaction between two agents who have to choose between
cooperation (Action 1) and defection (Action 2). If both players
decide to cooperate, each receives a \textquotedblleft
reward\textquotedblright\; payoff $R$ and if both players decide to
defect, then each receives a \textquotedblleft
punishment\textquotedblright\; payoff $P$. If one of the players
decides to defect and the other to cooperate, then the former gets a
payoff $T$ (often termed as the ``temptation'' to defect)  and the latter gets
the \textquotedblleft sucker's payoff\textquotedblright\; $S$. 

In PD the hierarchical relation between the different payoffs is
$T>R>P>S$. 
The only Nash equilibrium for this game is 
both agents choose defection (each receiving payoff $P$), as 
unilateral deviation by an agent
would yield a lower payoff ($S$) for her. Note that, mutual defection
is the only Nash solution even if the game is repeatedly played
between the players a finite number of times. 
However, it is easy to see that mutual cooperation would have resulted
in a higher payoff ($R$) for both agents.
This illustrates the apparently paradoxical aspect of the Nash solution
for PD where pursuit of self-interest by rational agents leads to a less
preferable outcome for all parties involved.
The failure on the part of the agents - who have been referred to as
``rational fools''~\cite{sen} - to see the obviously better
strategy is at the core of the dilemma and has important
implications for the social sciences, including economists' assumptions 
about the efficiency of markets~\cite{Morgan2012}. 
Further, experimental realizations of PD show that some
degree of cooperation is achieved when the game is played by human
subjects, which is at variance with the Nash
solution~\cite{Rapoport1965,andreoni,sally1995}.

In more general terms, PD raises questions about how cooperation can
emerge in a society of rational individuals pursuing their
self-interest~\cite{axelrod} and there have been several proposals to
address this issue. These have mostly been in the context of the
iterative PD (rather than the single-stage game that we are
considering here) and typically involve going beyond the standard
structure of the game, e.g., by introducing behavioral
rules such as direct or indirect reciprocity~\cite{Sigmund2010},
assuming informational asymmetry~\cite{kreps}, etc. 
By contrast, in the co-action solution, rational selfish agents
achieve non-zero levels of cooperation in the standard single-stage
PD, with the degree of
cooperation depending on the ratio of temptation $T$ to reward $R$.

\begin{figure}
\begin{center}
\includegraphics[width=0.69\linewidth]{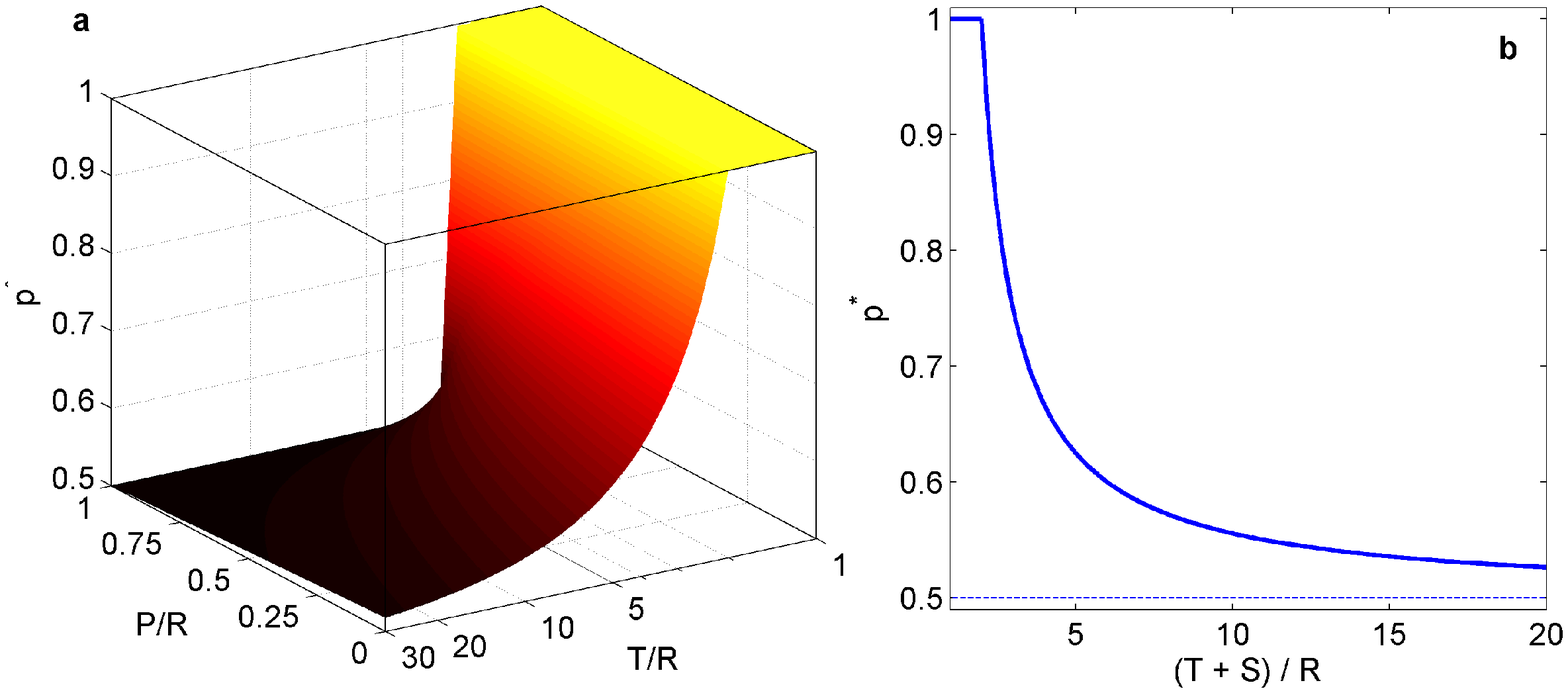}
\end{center}
\caption{The variation of the optimal strategy - probability of choosing Action
1, $p^{*}$ -
under the co-action solution concept for the games (a) Prisoner's
Dilemma (PD) and
(b) Chicken, as a function of the payoff matrix elements $T$, $P$, $R$ and
$S$. 
In both games, for low
values of $T$ (corresponding to temptation for defection in PD and
for being aggressive in Chicken), the
agents always opt for Action 1 (corresponding to cooperation in
PD and being docile in Chicken). However, as $T$ increases, agents 
opt for a mixed strategy, where Action 1 is chosen with decreasing
probability. In both cases, in the limit of very high $T$, the agent
strategy becomes fully random with 
the two actions being chosen with equal probability. Note that in PD,
the optimal strategy also has a very weak dependence on $P$ (corresponding to
punishment payoff for mutual defection).
}
\label{fig2}
\end{figure}

To obtain the co-action solution of PD, we use the formalism described
earlier with the value of the lowest payoff $S$ assumed 
to be zero without loss of generality. From
Eq.~\ref{co-action_payoff} and using the hierarchical relation among
the payoffs $T$, $R$ and $P$ for PD, it follows that
when $ T \leq 2R$, the optimal strategy for the agents is $p^* = 1$, i.e., 
both agents always cooperate.
On the other hand, when the temptation to defect $T > 2 R$, the
optimal strategy is a mixed one with the probability of cooperation
[Eq.~\ref{eq3}],
\begin{equation}
 p^* = \dfrac{T-2P}{2(T-R-P)},
\end{equation}
i.e., the agents
randomly choose between the available actions, defecting
with probability $1-p^*$. As temptation keeps increasing, the
probability of cooperation decreases and in the limit
$T \rightarrow \infty$, $p^* \rightarrow 1/2$, i.e., the agents
choose to cooperate or defect with equal probability, receiving an
expected payoff $W^* \rightarrow T/4$. Thus, unlike the Nash solution
of PD where cooperation is not possible, the co-action solution of the
game always allows a non-zero level of cooperation, with  
$1/2<p^*<1$ [Fig.~\ref{fig2}~(a)]. The co-action solution also differs
from the result expected based on the reasoning given by Rapoport and
Hofstadter~\cite{rapoport1966,hofstadter} - essentially a collective
rationality argument -
which suggests that rational agents will
always cooperate. 
To the best of our knowledge, co-action is the first solution concept which
allows probabilistic cooperation by the players in the single-stage
PD.

The existence of non-zero level of cooperation in the co-action
solution means that there is no longer any incompatibility
between the individual actions of rational agents trying to maximize their
payoffs and achieving the best possible collective outcome, thereby
resolving the ``dilemma'' in PD. The co-action concept may be used to
solve other games involving similar dilemmas such as traveler's
dilemma~\cite{basu1}.
It is of interest to note in this context that in the 
various experimental realizations of PD, the
level of cooperation observed is neither zero (as in the Nash
solution) nor complete - the average being about 50\% but with
significant variation across experiments~\cite{sally1995}.
While it is unclear if such realistic game conditions conform to the
idealized assumption of rational agents, the co-action solution
does provide a benchmark strategy for these situations.
%While the co-action solution may perhaps be too idealized to explain
%the results obtained under realistic conditions, it nevertheless
%provides a new benchmark strategy for such game situations. 
\subsection*{Chicken}
Chicken (also referred to as Snowdrift or
Hawk-Dove) is a two-person game that has been extensively investigated in the
context of the study of social interactions and evolutionary biology~\cite{maynard,rapoport}.
It represents a strategic interaction between two agents who have to
choose between being docile (Action 1) or being aggressive (Action
2). If both agents decide to be docile, they receive the payoff $R$,
while if one is docile when the other resorts to aggression, the
former - considered the ``loser'' - receives a lower payoff $S$ ($<R$)
and the latter - the ``winner'' - receives a
higher payoff $T$ ($>R$). 
However, the worst possible outcome corresponds to when both
players choose to be aggressive, presumably resulting in severe damage
to both, which is associated with the lowest payoff $P$. 
Thus, the hierarchical relation
between the different payoffs in Chicken is $T > R > S > P$. 
Note that it differs from PD in that the payoff $S$ is higher than
$P$. Therefore, an agent benefits by being 
aggressive only if the other is docile but is better off
being docile otherwise,
as the cost of mutual aggression is high.

The game has three Nash equilibria, of which two correspond to pure
strategies where one agent is docile while the other is aggressive.
The mixed strategy Nash
equilibrium 
$p_1^* = p_2^* =  S/(T+S-R)$ 
is given by Eq.~(\ref{nash}),
where it is assumed that the lowest of the possible payoffs $P$ is
zero [see Fig.~\ref{fig2}~(b)]. 
As in many other non-cooperative games with multiple Nash equilibria,
one has to invoke additional criteria (viz., equilibrium refinements
\cite{harsanyi88}) to decide which of these solutions will be selected
by the agents. In Chicken, a commonly used refinement concept is that
of evolutionarily stable strategy (ESS)~\cite{maynard} - an important
concept in evolutionary game theory~\cite{Hofbauer98} - 
which, in this game, gives the mixed strategy Nash equilibrium as the 
unique solution.

To obtain the co-action solution for Chicken, we note that 
under this solution concept, agents choose their actions so as to
optimize the payoff function Eq.~(\ref{co-action_payoff}).
Using the hierarchical relation of the payoffs for Chicken (assuming
the lowest payoff $P$ is zero
without loss of generality),
it is easy to see that for $2R \geq T+S$, $p^* = 1$ is the optimal
choice. 
On the other hand, when $2R < T+S$, agents choose to be docile with a
probability [Eq.~(\ref{eq3})],
\begin{equation}
 p^* = \dfrac{T+S}{2(T+S-R)}.
\end{equation}
Thus, for low values of $T$, both agents decide to be docile
(non-aggressive) always and avoid damaging each other, whereas, when the
stakes are high (for large $T$)
they randomly choose between the available actions, being docile
with probability $p^*$ and aggressive with probability $1 - p^*$. 
As in PD, in the limit of large $T$, i.e., $T\rightarrow\infty$, the
optimal strategy is $p^* \rightarrow 1/2$, where the agents choose to
be aggressive or docile with equal probability, receiving an expected
payoff $W^*\rightarrow T/4$.

It is instructive to compare the optimal strategy of the agents under
the different solution concepts when the stakes are very high. 
In the limit of $T\rightarrow\infty$, the ESS suggests that both
agents should resort to mutual aggression [i.e., $p_1^* = p_2^* \rightarrow
0$ which is evident from Eq.~(\ref{nash})]. This would result in both
agents suffering serious damage and receiving the lowest possible
payoff $P$. Compared to this, the co-action concept yields a
a significantly better outcome for both agents, as noted above. 
This difference is remarkable as the co-action solution shows that
``nice'' behavior among rational agents can occur
even in a highly competitive environment.
As in PD, experimental realizations of the single-stage game have
reported a significant
level (about $50\%$) of cooperative
behavior~\cite{Neugebauer2008}.

\subsection*{Stag Hunt}
The last of the two-person games we discuss here is the Stag Hunt
which is used to describe many social situations where cooperation is
required to achieve the best possible outcome~\cite{skyrms}.
The game
represents a strategic interaction between two agents who have to
choose between a high-risk strategy having potentially large reward,
viz., hunting for stag (Action~1) or a relatively low-risk, but poor-yield, 
strategy, viz., hunting for hare (Action~2). 
The agents can catch a stag (which is worth more than a hare) only if
they both opt for it, i.e., cooperate, thereby receiving the highest
payoff $R$. However, being unsure of what the other will do,
they may both choose the safer option of hunting hare, which can be done
alone, so that each receives a lower payoff $P$. However, if one agent
chooses to hunt stag while the other decides to hunt hare, the former
being unsuccessful in the hunt receives the lowest possible payoff
$S$, while the latter (who succeeds in catching hare) gets the payoff $T$. 
Thus, the hierarchical relation between the payoffs in
Stag Hunt is $R>T \geq P > S$.

As in Chicken, the game has three  Nash
equilibria, of which two correspond to pure strategies where both
agents opt for hunting stag or both choose to hunt hare. Note that
both strategies are
also evolutionarily stable, so that the ESS refinement, unlike in Chicken, 
does not yield a unique solution for this game.
The mixed strategy Nash equilibrium $p_1^* = p_2^* = P/(P+R-T)$ is 
given by Eq.~(\ref{nash}) where it is assumed that the lowest of the
possible payoffs $S$ is zero.

The co-action solution for Stag Hunt is obtained by noting that as
$R$ is greater than $T$, the payoff function [Eq.~(\ref{co-action_payoff})] 
increases monotonically in the interval $[0,1]$. Thus, the
co-action payoff $W = p^2(R+P-T) + p(T-2P)+P$ is optimized when $p^* =
1$, regardless of the values  of $R$, $T$ and $P$. Therefore, the
solution of the game under the co-action concept is unique, with both
agents opting to hunt stag, resulting in the best outcome for them.
It may be of interest to note that
experiments in single-stage Stag Hunt have reported that players tend
to choose to coordinate on the higher payoff outcome in the majority
of cases~\cite{skyrms,Battalio2001,Schmidt2003}.
Unlike in the previous two case-studies, there is no conflict
of interest among the agents playing Stag Hunt, who are instead trying
to coordinate their actions in the absence of any communication.
Thus, it can be viewed as a problem of equilibrium selection, with
the co-action solution corresponding to the better one.

\section*{Discussion}
\label{sec4}
In this paper we have shown that the conflict between pursuit of
individual self-interest and occurrence of collective outcomes that
are mutually beneficial in the context of social dilemmas
%non-cooperative games 
such as PD may only be an apparent one. The co-action concept presented
here resolves this conflict by making mutually consistent
assumptions about the behavior of rational agents.
The different games that are analyzed in detail here show that the
co-action solution concept leads to 
strategies that are relatively ``nicer'' and globally more efficient
compared to the standard Nash equilibrium concept.
In particular, it resolves the
dilemma in PD as the mutually beneficial action, viz., cooperation,
always has a significant probability ($\geq 1/2$) of being chosen by 
both agents. Similarly, co-action yields more cooperative outcomes in
the other games, i.e., agents playing Chicken resort to
non-aggressive strategies and agents achieve perfect coordination to
receive the highest possible payoff in Stag Hunt. Thus, this solution
concept reconciles 
the idea of individual self-interest pursued by rational agents with
the achievement of collective outcomes that are mutually beneficial,
even for single-stage games. 
%In addition, the co-action solutions for these games are also unique
%and do not require any additional refinement concepts.
While we do not claim that co-action is the only mechanism by which 
cooperation may originate and be maintained in nature, it certainly
shows that cooperation can evolve among selfish rational agents.
%solves the problem of 
%how cooperation originates and is maintained in nature, but it 
%certainly answers the question as to whether, in principle, selfish rational
%agents can achieve cooperation.
Note that our results do not depend on the specific definition of rationality
one uses, as long as the same definition applies to all agents.

For an $N$-player ($N>2$) game, if it can be
considered as the set of all pair-wise interactions between agents who
are symmetric in every respect, it is easy to see that the
optimal co-action strategy will be exactly the same as
that of the two-person game. 
The co-action solution concept can be generalized even to
cases where the symmetry assumption does not hold across all agents.
If the agents are aware
that some of the other agents are different from them, one can still
apply co-action within each cluster of agents (group) whose members consider
each other to be identical (i.e., the symmetry assumption holds).
For agents belonging to different groups, however, the payoffs are not
invariant under interchanging the identities of the players. Thus, the
symmetry of agents is broken across groups.
For a population of agents whose members can be considered as
belonging to two groups, one can treat the game as a two-player Nash-like
scenario where each ``player" is now a group of agents. 
However,
unlike the standard Nash setting where one cannot have a mixed
strategy as a stable Nash equilibrium, it is now possible for mixed
strategy equilibria to be stable~\cite{sasidevan14}.
In general, one can consider
a game with $N$ agents, clustered into $M$ symmetry groups, who have to choose
between two actions. Assuming that the
size of each group $i$ is $n_i$ ($\Sigma_i n_i = N$), the payoff for an
agent belonging to the $i$-th group 
is a polynomial of degree $n_i$ in $p_i$ ($i = 1, \ldots, M$),
where each $p_i$ is the probability of agents in that group to choose 
one of the actions. By contrast, the corresponding formulation of the
game in terms of Nash solution concept will involve $N$ variables with
the payoffs being linear in each of these variables.
Therefore, this defines a novel class of games between multiple
clusters of agents, with agents independently choosing the same
strategy as the other members of the cluster they belong to. 
The co-action results for such games may have potential implications
for multi-agent strategic interactions, as in the tragedy of
commons~\cite{commons}.

While the results discussed here are in the context of idealized
situations involving rational selfish agents, one may ask under what
conditions would the co-action framework apply in real life. As we
have outlined above, symmetry is a crucial ingredient for
co-action thinking to apply. Such symmetry is more likely to be
realized among members of a given community who share the 
same beliefs and a common identity.
It has indeed been observed that cooperation is more common
within an in-group than between agents belonging to different
groups~\cite{Balliet2014}.
The significant levels of cooperative behavior reported in
experimental realizations of social dilemmas (e.g., see
Refs.~\cite{Rapoport1965,sally1995} for PD and Ref.~\cite{Ledyard1995}
for its $N$-person generalization, i.e., the public goods game, 
Ref.~\cite{Neugebauer2008} for Chicken
and Refs.~\cite{skyrms,Battalio2001,Schmidt2003} 
for stag-hunt)
could, to some extent, be
explained by players ascribing to other players the same reasoning
process as themselves
and therefore resorting to co-action-like thinking.
%It has been explicitly shown that the cooperation in
%public-goods games experiments cannot be
%explained away as the result of errors or confusion among players, as
%about half of the observed cooperation is from players who choose to
%cooperate while aware of the problem of ``free-riding'', i.e., other
%players taking advantage of their cooperation~\cite{Andreoni1995}.
Experiments with human subjects playing PD have shown that the level of 
cooperation depends on the
actual values of payoffs and in general decreases with the ratio of
temptation for defection to reward for cooperation~\cite{sally1995} - in line with the
co-action solution. Also, players are known to employ
non-deterministic strategies in PD realizations~\cite{Cooper1996}, 
similar to what agents do in the
co-action equilibrium for sufficiently high temptation.
Game situations that allow ``cheap talk'' (i.e., communication between
agents that does not directly affect payoff)~\cite{Farrell1996} which
presumably allow players to affirm shared set of values - and thereby
promote co-action thinking -
have been shown to increase the level of cooperation in
experiments~\cite{Crawford1998}.
In other experimental realizations, where players in a public goods
game indicated their
preferred contributions for different average levels of
contribution by other group members, about half the players were observed
to match what the others would do~\cite{Fischbacher2001}. 
Such ``conditional cooperation'' could be an illustration of the
symmetry considerations that players might engage in. 
Also, the co-action framework could provide a natural setting for the
emergence of tag-based cooperation schemes among ``sufficiently
similar'' agents~\cite{Riolo2001}. 
%In particular, it may hold in situations where the agents are
%``sufficiently similar'', thereby providing a natural setting for the
%emergence of tag-based
%cooperation schemes~\cite{Riolo2001}.
The idea of ``social projection''~\cite{Krueger2012} provides 
yet another instance where such
considerations may be relevant.
We note that there have been other approaches towards
explaining cooperation in social dilemmas based on symmetry of the
game situation, e.g., a recent model in which agents decide their strategies
based on their most optimistic forecast about how the game would be
played if they formed coalitions~\cite{Capraro2013}.

In this paper we have focused on single-stage games but the co-action
concept discussed here applies also to repeated games where information
about the choices made by agents in the past are used to decide their
future action. In this situation, the co-action solution developed in the context of 
single-stage games is applied at each iteration, with the past
actions of agents used to define the different symmetry
groups~\cite{sasidevan}.
This is inherently a dynamical process, as
the membership of these groups can evolve in time.
For example, in iterative PD with $N$ agents having 
memory of the choices made in the previous iteration, 
all agents who made the same decision in the last round will belong to
the same symmetry group and will behave identically. 
The resulting solution can allow coexistence
of cooperators and defectors in the game, which we will discuss in a future 
publication.

To conclude, we have introduced here a solution framework for
non-cooperative games that
%which makes use of the symmetry between rational agents.
%The resulting co-action equilibrium for a game have properties
%radically different from that of the corresponding Nash equilibrium,
%the conventional solution employed for such games. In particular, the
%co-action concept 
resolves the apparent conflict between rationality
of individual agents and globally efficient outcomes.
%in games such as PD.
It suggests that cooperation can evolve in nature as the rational
outcome even with
selfish agents, without having to take recourse to additional
mechanisms for promoting it.
In practice, the co-action and Nash solutions could represent two
extreme benchmark strategies for non-cooperative games, the latter
applying when the agents cannot be considered to be ``sufficiently
similar''. While we do not address
here the question of which concept is more appropriate for a given
situation, it is conceivable that agent behavior in reality may be
described by a strategy between these two extremes and can potentially be
represented by a combination of them. 
Although we have discussed co-action in the context of the evolution of
cooperation among rational agents, the concept is far more general and
could provide a mechanism for understanding strategic interactions
across groups of sufficiently 
similar agents in many different settings.
%The concept of co-action is far more general than the context of
%evolution of cooperation discussed here and provides a mechanism for
%understanding strategic interactions across groups of sufficiently
%similar agents in many different settings.
%In general, the concept of
%co-action provides a mechanism for addressing the question of 
%evolution of cooperation among sufficiently similar agents in
%nature.
\section*{Acknowledgements}
This work was partially supported by the IMSc Econophysics project
funded by the Department of Atomic Energy, Government of India. We
thank Deepak Dhar for useful discussions and Shakti N. Menon for useful
comments on the manuscript.


\begin{thebibliography}{99}
\bibitem{Morgenstern44} Morgenstern, O. \& Von Neumann,  J. 
{\em Theory of games and economic behavior} (Princeton University
Press,
Princeton, 1944).

\bibitem{colman}  Colman, A. M. {\em Game theory and its
applications
in the social and biological sciences} (Routledge, New York, 1999).

\bibitem{hargreaves} Hargreaves Heap, S. P. \& Varoufakis, Y. 
{\em Game theory: A critical text} (Routledge, London, 2002).

\bibitem{osborne}  Osborne, M. J. \&  Rubenstein, A. {\em  A course in
game theory} (MIT Press, Cambridge, Mass. 1994).

\bibitem{Nash1950} Nash, J. F. Equilibrium points in n-person
games. {\em Proc. Natl. Acad. Sci.
USA.} {\bf 36(1),} 48-49 (1950).

%Equilibrium points in n-person games. PNAS 36.1 (1950): 48-49.
\bibitem{Holt2004}  Holt, C. A.\& Roth, A. E.  The Nash equilibrium: A
perspective. {\em Proc. Natl. Acad. Sci.
USA.} {\bf 101(12),} 3999-4002 (2004).
%The Nash equilibrium: A perspective. PNAS 101.12 (2004): 3999-4002.
\bibitem{harsanyi88} Harsanyi, J. C. \& Selten, R.  {\em A General
Theory of
Equilibrium Selection in Games} (MIT Press, Cambridge, 2003).

\bibitem{binmore} Carlsson, H. \& van Damme, E.  Equilibrium selection
in stag hunt games in 
{\em Frontiers of Game Theory} (eds Binmore, K. G., Kirman, A. P. \& Tani, P.) (MIT
Press, Cambridge MA, 1993).

\bibitem{Rapoport1965} Rapoport, A. \& Chammah, A. M.  {\em Prisoners
Dilemma} (Univ of Michigan Press, Ann Arbor, MI, 1965).

\bibitem{basu1} Basu, K.  The traveler's dilemma: Paradoxes of
rationality in game theory. {\em The American Economic Review}
{\bf 84(2),} 391-395 (1994).

\bibitem{basu2} Basu, K.  The traveler's dilemma. {\em Scientific
American Magazine} {\bf 296(6),} 90-95 (2007).

\bibitem{andreoni} Andreoni, J. \& Miller, J.  Rational cooperation
in the finitely repeated prisoner's dilemma: Experimental evidence.
{\em The Economic Journal} {\bf 103,} 570-585 (1993).

\bibitem{Kollock98} Kollock, P.  Social dilemmas: The anatomy of
cooperation. {\em Annual Review of Sociology} {\bf 24,} 183-214 (1998).

\bibitem{axelrod} Axelrod, R.  {\em The Evolution of Cooperation}
(Basic Books, New York, 1984).

\bibitem{sasidevan} Sasidevan, V. \& Dhar, D.  Strategy switches and
co-action equilibria in a minority game. {\em Physica A: Statistical
Mechanics and its Applications} {\bf 402,} 306-317 (2014).

%Social dilemmas: The anatomy of cooperation
%\bibitem{stanford}
%http://plato.stanford.edu/entries/prisoner-dilemma/
\bibitem{rapoport1966}
Rapoport, A. {\em Two-person game theory: The essential ideas} 
(University of Michigan Press, Ann Arbor, 1966).

\bibitem{hofstadter} Hofstadter, D.  The calculus of cooperation
is tested through a lottery. {\em
Scientific American} {\bf 248(6),} 14-28 (1983). Reprinted in: Hofstadter, D. 
{\em Metamagical Themas} 737-755 (Basic Books, New York, 1985).

\bibitem{Binmore1994}
Binmore, K. G. {\em Game theory and the social contract. Vol 1. Playing
fair} (MIT Press, Cambridge, 1994).

\bibitem{McMahon2001} McMahon, C.  {\em Collective Rationality and
Collective Reasoning} (Cambridge University Press, Cambridge, 2001).

\bibitem{Archetti2012} Archetti, M. \& Scheuring, I. Review: Game
theory of public goods in one-shot social dilemmas
without assortment. {\em Journal of Theoretical Biology} {\bf 299,}
9-20 (2012).

%\bibitem{davis1977}  Davis, L. H.  Prisoners, paradox, and
%rationality. {\em American Philosophical Quarterly} {\bf 14,} 319-327 (1977).

%\bibitem{rapoport1967} Rapoport, A. Escape from paradox. {\em Scientific
%American} 50-56 (1967).

%\bibitem{howard1971} Howard, N. Paradoxes of rationality: Theory of metagames and political
%behavior (MIT Press) (1971).

\bibitem{rapoport} Rapoport, A. \& Chammah, A. M.  The game of chicken.
{\em American Behavioral Scientist} {\bf 10(3),} 10-28 (1966).

\bibitem{nowak2004} Nowak, M. A. \&  Sigmund, K.  Evolutionary dynamics
of biological games. {\em Science}  {\bf 303(5659),} 793-799 (2004).

\bibitem{sen} Sen, A.  Rational fools: A critique of the
behavioral foundations of economic theory. {\em Philosophy and Public
Affairs} {\bf 6,} 317-344 (1977).

\bibitem{Morgan2012} Morgan, M. S.  {\em The World in the Model}
(Cambridge University Press, New York, 2012). 

\bibitem{sally1995} Sally, D.  Conversation and Cooperation in
Social Dilemmas A Meta-Analysis of Experiments from 1958 to 1992. {\em
Rationality and Society} {\bf 7(1),} 58-92 (1995).

\bibitem{Sigmund2010} Sigmund, K.  {\em The Calculus of
Selfishness} (Princeton University Press, Princeton, NJ, 2010).

\bibitem{kreps} Kreps, D., Milgrom, P., Roberts, J. \&  Wilson, R. 
Rational cooperation in the finitely repeated prisoners' dilemma. {\em
Journal of Economic Theory} {\bf 27(2),} 245-252 (1982).

\bibitem{maynard} Smith, J. M. \& Price, G. R.  The Logic of Animal
Conflict.  {\em Nature} {\bf 246,} 15-18 (1973).

\bibitem{Hofbauer98} Hofbauer, J. \& Sigmund, K.  {\em Evolutionary
Games and Population Dynamics} (Cambridge University Press,
Cambridge, 1998).

\bibitem{Neugebauer2008} Neugebauer, T., Poulsen, A. \& Schram, A. Fairness and reciprocity in the Hawk–Dove game. {\em Journal of Economic Behavior \& Organization} {\bf 66,} 243-250 (2008).


\bibitem{skyrms} Skyrms, B.  {\em The Stag Hunt and the Evolution
of Social Structure} (Cambridge University Press, Cambridge, MA, 2004).
%\bibitem{nowak} Nowak M. A., Page K. M. and Sigmund K. (2000).
%"Fairness Versus Reason in the Ultimatum Game". Science 289 (5485):
%1773–1775.
\bibitem{Battalio2001} Battalio, R., Samelson, L. \& Huyck, J. V. Optimization incentives and coordination failure in laboratory Stag Hunt games. {\em Econometrica} {\bf 69 (3),} 749-764 (2001).

\bibitem{Schmidt2003} Schimdt, D., Shupp, R., Walker, J. M. \& Ostrom, E. Playing safe in coordination games:
the roles of risk dominance, payoff dominance,
and history of play. {\em Games and Economic Behavior} {\bf 42,} 281–299 (2003).

\bibitem{sasidevan14} 
Stability of a
solution can be defined with respect to small changes in strategy
parameters. We discuss this in 
%Ref.~\cite{sasidevan15}
Sasidevan, V. \& Sinha, S.  A Dynamical view of
different solution paradigms in two-person symmetric games: Nash
versus co-action equilibria in {\em  Econophysics and Data Driven
Modelling of Market Dynamics} (eds Abergel, F. et al.) (Springer,
Milan, 2015), where Nash and 
co-action equilibria for two-person games are viewed in a dynamical
systems framework.

\bibitem{commons}
Hardin, G. The tragedy of the commons. {\em Science} {\bf 162,}
1243-1248 (1968).

%\bibitem{dorrough2015} Dorrough, A. R., Glockner, A. Hellmann, D. M. \& Ebert, I. The development of ingroup
%favoritism in repeated social dilemmas. {\bf 6,} 476 (2015).
\bibitem{Balliet2014}
Balliet, D., Wu, J. \& De Dreu, C. K. W. Ingroup favoritism
in cooperation: a meta-analysis. {\em Psychol. Bull.} {\bf 140,}
1556–1581 (2014). 
%doi:10.1037/a0037737

\bibitem{Ledyard1995}
Ledyard, J. O. Public goods: A survey of experimental resarch in
{\em Handbook of Experimental Economics} (eds Kagel, J. \&
Roth, A.) (Princeton University
Press, Princeton NJ, 1995).

%\bibitem{Andreoni1995}
%Andreoni, J. Cooperation in Public-Goods Experiments: Kindness or
%Confusion~? {\em American Economic Review} {\bf 85,} 891-904 (1995).

\bibitem{Cooper1996} Cooper, R., Dejong, D. V., Forsythe, R \& Ross, T. W. Cooperation without reputation: Experimental evidence from Prisoner’s Dilemma games. {\em Games and Economic Behavior} {\bf 12,} 187-218 (1996).

\bibitem{Farrell1996}
Farrell, J. \& Rabin M. Cheap talk. {\em Journal of Economic
Perspectives} {\bf 10,} 103-118 (1996).

\bibitem{Crawford1998}
Crawford, V. A survey of experiments on communication via cheap talk.
{\em Journal of Economic Theory} {\bf 78,} 286-298 (1998).

\bibitem{Fischbacher2001} Fischbacher, U., Gachter, S. \& Fehr, E. Are people conditionally cooperative? Evidence from a public
goods experiment. {\em Economics Letters} {\bf 71,}  397–404 (2001).

\bibitem{Riolo2001}
Riolo, R. L., Cohen, M. D. \& Axelrod, R.  Evolution of cooperation
without reciprocity. {\em Nature} {\bf 414}: 441-443 (2001).

\bibitem{Krueger2012} Krueger, J. I., DiDonato, T. E. \& Freestone, D.
Social projection can solve social dilemmas.
{\em Psychological Inquiry} {\bf 23,} 1-27 (2012).

\bibitem{Capraro2013} Capraro, V. A model of human cooperation in social dilemmas. {\em Plos One} {\bf 8(8),} e72427 (2013).


%\bibitem{upcoming} Sasidevan V, Sinha S (2015) in preparation.
%\bibitem{sasidevan15}
%Sasidevan, V. \& Sinha, S.  A Dynamical view of
%different solution paradigms in two-person symmetric games: Nash
%versus co-action equilibria in {\em  Econophysics and Data Driven
%Modelling of Market Dynamics} (eds Abergel, F. et al.) (Springer, Milan, 2015).

%\bibitem{downs} Downs, A. (1957), An Economic Theory of Democracy,
%Harper and Row, N.Y., 1957.
%\bibitem{stanford}
%http://plato.stanford.edu/entries/prisoner-dilemma/.

%\bibitem{dhar} D. Dhar, V. Sasidevan and Bikas K Chakrabarti, Physica
%A {\bf 390}, 3477 (2011).
%\bibitem{fundenberg} D. Fudenberg and J. Tirole, {\it Game theory},
%(MIT Press, Cambridge, MA 1991).

\end{thebibliography}
\end{document}